\documentclass[11pt]{article}
\usepackage{epsfig,amscd,amssymb}
\usepackage{graphicx}
\usepackage{latexsym}
\begin{document}

%
%
%
%
\title{Non-local space-time supersymmetry on the lattice}%

\author{Xiao Yang and Paul Fendley\\
Department of Physics\\
University of Virginia\\
Charlottesville, VA 22904-4714\\
{\tt xy9n@virginia.edu, fendley@virginia.edu}
}
\maketitle

\begin{abstract}
We show that several well-known one-dimensional quantum systems
possess a hidden non-local supersymmetry. The simplest example is the
open XXZ spin chain with $\Delta=-\frac{1}{2}$. We use the supersymmetry to
place lower bounds on the ground state energy with various boundary conditions.
For an odd number of sites in the periodic chain, and with a particular
boundary magnetic field in the open chain, we can derive 
the ground-state energy exactly. The supersymmetry thus explains why
it is possible to solve the Bethe equations for the ground state in
these cases. We also show that a similar space-time supersymmetry holds
for the t-J model at its integrable ferromagnetic point, where the
space-time supersymmetry and the Hamiltonian it yields
coexist with a global $u(1|2)$ graded Lie
algebra symmetry. Possible generalizations to other algebras are
discussed. 

\end{abstract}

\section{Introduction}

In studying strongly correlated systems, one cannot rely on
conventional perturbation theory.  It is therefore useful to
explore the symmetries of such models in depth.  Supersymmetry is
a fairly generic term meaning that some of the symmetry generators
are fermionic, and so obey anticommutation
relations.  ``Space-time'' supersymmetry is quite special, because
the Hamiltonian not only commutes with the symmetry
generators, but is a part of the symmetry algebra -- it appears in an
anticommutator of fermionic generators. A number of important properties follow
from this fact. For example, all the energies obey $E\ge 0$.

In this paper we mainly study ${\cal N}=2$ supersymmetry
\cite{Witten}.  Here there are two hermitian-conjugate supercharges,
which we denote $Q$ and $Q^{\dagger}$. The charges are nilpotent,
which means they obey $Q^2 = (Q^\dagger)^2 = 0$. Their anticommutation
relation yields the Hamiltonian:
\begin{equation} \label{eq:anticom} \{Q, Q^{\dagger}\} = H 
\end{equation}
An additional bosonic symmetry generator is the fermion
number $F$, which obeys
\begin{equation}
[F,Q] = - Q\ ,\qquad \quad [F,Q^\dagger]=Q^\dagger
\end{equation}
Because the Hamiltonian is $QQ^\dagger +Q^\dagger Q$, its eigenvalues
$E$ cannot be negative. Any states with $E=0$ are therefore ground
states, and must be annihilated by both $Q$ and $Q^\dagger$.  States
with $E>0$ form doublets under the supersymmetry.  The two states in
the doublet have opposite values of $(-1)^F$, so no states with $E>0$
contribute to the Witten index $W=\hbox{tr}\,(-1)^F e^{-\beta H}$. Thus $W$
is precisely the number of bosonic ground states minus the number of
fermionic ground states, independent of $\beta$ and $H$.

A number of lattice models result in supersymmetric Lorentz-invariant
field theories in the continuum limit \cite{Saleur}. However, only a
few lattice models are known where the space-time supersymmetry
defined above is present explicitly on the lattice
\cite{Fendley02,Fendley03}.  The simplest model discussed in
\cite{Fendley02,Fendley03} consists of spinless fermions with a
hard-core interaction, with in particular the restriction that they
cannot be on the same or on adjacent sites.  It was shown in that the
energy levels (up to an overall shift) of this theory are the same as
that of the XXZ spin chain at a particular value of the anisotropy
$\Delta=-1/2$ and with particular twisted boundary conditions.  This
spin chain here is known to yield a supersymmetric field theory in its
continuum limit, so it is not shocking that there is a fermion model
in the same universality class which has explicit supersymmetry on the
lattice. What is somewhat surprising is that the correspondence between
the explicitly-supersymmetric model and the spin chain persists even
on the lattice.

This suggests that this XXZ chain is supersymmetric in its own right,
so that its Hamiltonian can be written in the form
(\ref{eq:anticom}). There are no fermions in the XXZ model, so the
supercharges must necessarily be non-local combinations of the
spins. Such a construction of a fermionic operator from bosonic ones
in one dimension is familiar from the Jordan-Wigner transformation of
the XXZ model \cite{LMS}. 

The purpose of this paper is to show that the XXZ chains at $\Delta =
-1/2$ are indeed supersymmetric, and to construct their
supercharges. One consequence of this result is that this
automatically yields the ground state energy.  The reason is that the
supersymmetry requires that the Hamiltonian be of the form $Q
Q^\dagger + Q^\dagger Q + E_0$, where $E_0$ is a known
(size-dependent) constant. This already means the ground-state energy
is bounded from below at $E_0$, but in some cases we will derive that
the ground-state energy is precisely $E_0$.  Analogous results
are known as a result of elaborate Bethe ansatz computations
\cite{Bax8v} and by utilizing the Temperley-Lieb algebra \cite{Gier},
but our result gives this simply and directly.  Knowing
the ground-state energy exactly in a system solvable by the Bethe
ansatz is quite useful, because then the ground-state wave function
can be characterized in terms of the roots of a single polynomial
equation \cite{FSZ,Stroganov,twistedXXZ}.

Amusingly enough, precisely at this value of $\Delta$, the chain is
experimentally realizable by putting a spin-1 chain in a Haldane gap
phase in a magnetic field tuned to make one of the spin-triplet
excitations degenerate with the ground state \cite{Diederix}. The
magnetic field breaks the $SU(2)$ symmetry, this two-state system
becomes an antiferromagnetic XXZ chain at some value of $\Delta$; it
is not difficult to show that this is precisely $\Delta=-1/2$. Thus
our result provides an experimental realization of supersymmetry!

We also extend our results in several ways. We show that the t-J model
at its integrable ferromagnetic point has an explicit supersymmetry,
as suggested by the results of \cite{Fendley03}. We also present a
Hamiltonian which commutes with supercharges $R$ and $R^\dagger$ which
obey $R^4=(R^\dagger)^4=0$. This model resembles the XXZ chain, but
does not seem to be equivalent.

\section{Supersymmetry in the XXZ model}

In \cite{Fendley03}, a series of space-time supersymmetric lattice
models of spinless fermions $M_k$ was constructed. In $M_k$, the
Hilbert space is restricted so that no more than $k$ consecutive sites
can be occupied.  It was found \cite{Fendley03} that for $M_{1}$
model, if an edge between two empty sites is mapped to an up spin, and
an occupied site together with its two adjacent edges is mapped to a
down spin, then it is closely related to the XXZ model at a particular
coupling. However, the mapping is not always one-to-one,
and requires care with the boundary conditions. 
Thus although this result strongly suggests the
supersymmetry appears directly in the XXZ model, it does not prove
it. In this section we show how the supersymmetry indeed appears
directly.

The XXZ model is a generalization of the Heisenberg model.
The Hamiltonian acts on quantum spins
$\vec{S}_j$ in the spin-$1/2$ representation of $SU(2)$
on each site $j$:
\begin{equation}
\label{eq:xxz-Hamiltonian} H_{XXZ} = -h (S^z_1 + S^z_L)
-2\sum^{L}_{j=1}(S^{x}_{j}S^{x}_{j+1} +
S^{y}_{j}S^{y}_{j+1} + \Delta S^{z}_{j} S^{z}_{j+1})
\end{equation}
where we have allowed for a boundary magnetic field $h$.  For
$\Delta=1$ and $\Delta=-1$, one recovers the $SU(2)$-symmetric
ferromagnetic and antiferromagnetic Heisenberg models
respectively. When $|\Delta|\ne 1$, the $SU(2)$ symmetry is broken
down to $U(1)$; the conserved charge is say the number of down spins
$n_{\downarrow}$. 
In this paper we focus mainly on the case $\Delta =-1/2$.

\subsection{Open chain}

The easiest way to describe the supersymmetry
is to utilize a fermionic representation of the spins.
We introduce two species
of fermions $f_{j\downarrow}$ and $f_{j\uparrow}$, and require that
there be one fermion of either species at each site $j$. 
Spin operators in the spin-$1/2$ representation can be
written as
\begin{equation}
\label{eq:fermionrp}
\vec{S}_{j}=f^{\dagger\alpha}_{j}\vec{\sigma}_{\alpha}^{\beta}f_{j\beta}
\end{equation}
where the $\vec{\sigma}$ are Pauli matrices, and
$\alpha$ and $\beta$ are spin indices which can take value $\uparrow$ and
$\downarrow$. In terms of raising and lowering operators, we have
$$S^{+}_j= f^\dagger_{j\uparrow} f_{j\downarrow}, \qquad\quad
S^{-}_j= f^\dagger_{j\downarrow} f_{j\uparrow}$$

Using the fermionic representation, we can find supercharges $Q$ and
$Q^\dagger$ which commute with the Hamiltonian
(\ref{eq:xxz-Hamiltonian}) when $\Delta=-1/2$.  To do this, we need to
define the ``shift'' operator $A^{R\dagger}_{j}$, which moves all the
fermions on sites $k>j$ to the right by one. It thus increases the number
of sites in the system by one, while leaving an unoccupied site at $j+1$. 
The supercharge $Q$ is then given by
$$Q =\sum_{j=1}^L Q_{j},$$
where
\begin{equation}
\label{eq:xxz-charge}
Q_{j} =f^{\dagger}_{j\uparrow}f^{\dagger}_{j+1,\uparrow}f_{j\downarrow}
A^{R\dagger}_{j}
\end{equation}
The supercharge is non-local because of the shift.  Simply speaking,
the supercharge $Q$ converts a down spin at site $j$ to two up spins
at $j$ and $j+1$. As a result, the total number of fermions as well as
the number of sites is increased by one.  To be precise, we define the
Hilbert space to have $L+1$ sites with the $L+1$th site empty. Then
$Q_{L+1}=0$.  The fermion-number generator $F$ in the superalgebra
merely corresponds to the generator of $U(1)$ symmetry in the XXZ
model: 
$$F=\sum_{j=1}^L f_{j,\downarrow}^{\dagger} f_{j,\downarrow}^{}$$

It is simple to verify that $Q^2=0$ by using the fermion anticommutation
relations to verify that
$Q_j Q_k + Q_{k+1} Q_j=0$ for $j\le k$ and that $Q_j^2=0$.  
It is instructive to study how these operators change spin
configurations. 
Let the Greek letters $\alpha,\beta,\dots$ represent
up or down spins, so that the
spin configuration $\big|\alpha\;\beta\;...\;\zeta\rangle$
is the state
$f^{\dagger}_{1\alpha}f^{\dagger}_{2\beta}...f^{\dagger}_{L\zeta}
\big|0\rangle$,
where $\big|0\rangle$ is the vacuum state. 
First, let's check the anticommutator 
$2Q^2=\{Q, Q\} =\sum_{i,j}\{Q_{i}, Q_{j}\}=0$. 
Let's first see how $Q_i Q_j$ for $i<j$ acts. This automatically
vanishes except on configurations which have down
spins at the $i$th and the $j$th sites.
Then $Q_j$ acts on such a configuration $\big|S\rangle$ as
$$Q_j\big|S\rangle = Q_j\big|\dots\;\alpha\,\downarrow\,\beta \;\dots\; \eta\,\downarrow\, \zeta\;\dots\rangle
= (-1)^{j-1}\big|\dots\;\alpha\,\downarrow\;\beta \;\dots\; \eta\,\uparrow\,\uparrow\,\zeta\;\dots\rangle$$
where the factor $(-1)^{j-1}$ arises because
the 3 fermionic operators
$f^{\dagger}_{j,\uparrow}f^{\dagger}_{j+1,\uparrow}f_{j,\downarrow}$
are moved over $j-1$ fermionic sites. 
Now acting with $Q_i$ gives
$$Q_iQ_j\big|S\rangle =
(-1)^{i+j-2}\big|\dots\;\alpha\,\uparrow\,\uparrow\;\beta \;\dots\; \eta\,\uparrow\,\uparrow\,\zeta\;\dots\rangle$$
%
If we now compute $Q_{j+1}Q_i$ on the same configuration, we get the same
final state, except multiplied by $(-1)^{i+j-1}$.
Thus these two contributions to $Q^2$ cancel. 
The only terms surviving these cancellations, $Q_{i+1}Q_{i}$ and $Q_i^2$,
individually vanish. Thus we have proved that $Q$ is
nilpotent. 

Because $Q^2=(Q^\dagger)^2=0$, a Hamiltonian constructed via $H=\{ Q,
Q^\dagger\}$ commutes with the charges, and so is supersymmetric.
Even though $Q$ increases the number of spins by one, 
$Q^\dagger$ decreases them by one, so that the Hamiltonian preserves
the number of spins.  We have
$$H = \sum_{i=1}^{L-1} \sum_{j=1}^{L-1} Q_i Q_j^{\dagger} +
\sum_{j=1}^{L} \sum_{i=1}^{L} Q_j^{\dagger} Q_i $$
where we use the fact that $Q_{L+1}=0$ and $Q^\dagger_L=0$, 
when acting on our Hilbert space with $L$ fermions and the
$(L+1)$th site empty. 
The only states on which $Q_{i}Q^{\dagger}_{j}$ for $i<j$  
are non-vanishing are of the form
$$\big|S\rangle=
\big|\dots\;\alpha\,\downarrow\,\beta \;\dots\; \eta\,\uparrow\,\uparrow\,\zeta\;\dots\rangle$$
Acting with $Q_iQ_j^{\dagger}$ gives
$$Q_iQ_j^{\dagger}\big|S\rangle=(-1)^{i+j-2}
\big|\dots\;\alpha\,\uparrow\,\uparrow\,\beta \;\dots\; \eta\,\downarrow\,\zeta\;\dots\rangle$$
We also have 
$$Q_{j+1}^{\dagger}Q_i\big|S\rangle=(-1)^{i+j-1}
\big|\dots\;\alpha\,\uparrow\,\uparrow\,\beta \;\dots\; \eta\,\downarrow\,\zeta\;\dots\rangle$$
so that these two terms cancel each other.
Similarly, $Q^{}_{i}Q_{j}^{\dagger}$ for $i>j$ cancels with
$Q^{\dagger}_{j}Q^{}_{i+1}$.
Such cancellations get rid of almost all the terms, leaving only
\begin{equation}
H = \sum_{i=1}^{L-1} \left[Q^{\dagger}_{i+1}Q^{}_{i} + Q^{\dagger}_{i} 
Q^{}_{i+1} + Q_{i}^{} Q^{\dagger}_i \right] + \sum_{i=1}^L\left[
Q^{\dagger}_{i}Q^{}_{i}  \right]
\label{ham}
\end{equation}

We can now rewrite this Hamiltonian (\ref{ham}) in terms of the spins.
The first term acts as
$$
Q^{\dagger}_{i+1}Q_{i}
\big|\dots\;\alpha\,\downarrow\,\uparrow\,\zeta\;\dots\rangle=
(-1)^{2i-1}\big|\dots\;\alpha\,\uparrow\,\downarrow\,\zeta\;\dots\rangle
$$
This means that
$$Q^{\dagger}_{i+1}Q_{i} = - S^+_{i}S^-_{i+1}$$
The second term in (\ref{ham}) is likewise
$Q^{\dagger}_{i}Q_{i+1} = - S^+_{i+1}S^-_{i}$.
The last term simply counts the number of down spins:
$Q^{\dagger}_{i}Q^{}_{i}= f^{\dagger}_{i\downarrow} f_{i\downarrow} \equiv n_{i\downarrow}$. Finally, the third counts the number of adjacent up spins:
$$Q^{}_{i}Q^{\dagger}_{i}= n_{i\uparrow}n_{i+1,\uparrow}$$
These last two terms can be rewritten in terms of $S^z$ by 
noting that on any site occupied by a fermion, we have
$n_{j\downarrow} + n_{j\uparrow} =1$, and 
$$n_{j\uparrow} = \frac{1}{2} + S^z_j.$$
Putting this all together means that the Hamiltonian generated by
the supercharge (\ref{eq:xxz-charge}) is
\begin{equation}
\label{eq:xxzsusyh}
H = \frac{3}{4}L - \frac{1}{4}-\frac{1}{2} S^z_{1} 
- \frac{1}{2} S^z_{L}
-\sum_{j=1}^{L-1}\left[S^{+}_{j}S^{-}_{j+1} + S^{-}_{j}S^{+}_{j+1} 
-S^z_{j}S^z_{j+1}\right]
\end{equation}
Comparing with (\ref{eq:xxz-Hamiltonian}) we have
\begin{equation}
H_{XXZ} = H - \frac{3L-1}{4}
\end{equation}
where $\Delta=-1/2$ and the boundary magnetic field $h=1/2$.

The fact that the eigenvalues of $H$ are non-negative means that the
spectrum of $H_{XXZ}$ is bounded from below by $-(3L-1)/4$. By
using the supersymmetry, we can prove that this is in fact the
ground-state energy.  The first step is to compute the Witten index
$W=\hbox{tr}\,(-1)^F e^{-\beta H}$ \cite{Witten}. As noted in the
introduction, under the action of the supersymmetry generators, all
states other than the ground state form pairs with the same eigenvalue
of $H$ and opposite values of $(-1)^F$. Thus $W$ only receives
non-zero contributions from the ground states with zero eigenvalue of
$H$.  This means $W$ is independent of $\beta$ and $H$: it is just a
property of the Hilbert space of states. If $W$ is non-zero, then we
know that there is at least one ground state in the XXZ chain with
energy precisely $-(3L-1)/4$.

The subtlety in our particular case is that while $H$ preserves the
number of sites $L$, $Q$ effectively increases $L$ by one, while
$Q^\dagger$ decreases it. However, note that $H_{XXZ}$ also preserves
the number of down spins $n_{\downarrow}$, while $Q$ decreases
$n_\downarrow$ by 1. Since $Q$ and $Q^\dagger$ preserve the
combination $L + n_{\downarrow}$, it is useful to define the Hilbert
spaces $H_{L,n_\downarrow}$.  Then $Q$ takes a state in
$H_{L,n_\downarrow}$ to $H_{L+1,n_\downarrow-1}$, and $Q^\dagger$
vice-versa.  We can then define the Witten index $W_N$ to have the trace
taken over all states in the Hilbert spaces with fixed $N\equiv L +
n_{\downarrow}+1$. A non-zero value means that there is at least one
$E=0$ ground state in the Hilbert spaces with fixed $N$.

$W_N$ has already been computed for the case at hand, because this
reduces to the computation done for an open fermionic chain in
\cite{Fendley03}.  The $M_{1}$ model describes spinless fermions
hopping on an $N$-site chain, with a further constraint that no
adjacent sites can be simultaneously occupied. Thus there are two
states per site (empty and occupied) in the $M_1$ model, just like in
the XXZ model.  To match states in the two models with open boundary
conditions, we consider configurations in the $M_1$ model with the
first and last site empty.  Then each edge between empty sites in
$M_1$ is mapped to an up spin, while each fermion is mapped to a down
spin \cite{Fendley03}.  Thus the number of fermions $f$ in $M_1$ is
the number of down spins $n_{\downarrow}$, while the length $N$ of the
open $M_1$ chain is related to $L$ by $N=L+n_{\downarrow}+1$. The
Witten index $W_N$ in the two cases is then identical, but in $M_1$ it
is easy to phrase: it is the sum over all allowed configurations on a
chain of length $N-2$ (or equivalently length $N$ with first and last
sites unoccupied) with weight $(-1)^F$. This is easily computed using
a classical one-dimensional transfer matrix, in the same fashion as
one solves the one-dimensional Ising model.  One finds that $W_N=0$
for $N=3j$, $j$ integer, while $W_N=(-1)^{[N/3]}$ otherwise, where $[N/3]$
denotes the largest integer which is less than $N/3$ \cite{Fendley03}.

We have thus shown that for $N\ne 3j$, there is a state where $H$ has
a zero eigenvalue, so that $H_{XXZ}$ has eigenvalue $-(3L-1)/4$.  We
now need to translate this into a statement depending on $L$.  This
requires finding for a given $N$, the value of $n_{\downarrow}$ for
the ground state, hence yielding $L$.  Again, this question can be
phrased in terms of the open chain discussed in \cite{Fendley02}. The
key observation is that only ground states are annihilated by both $Q$
and $Q^\dagger$. Denoting $n_0$ to be the value of
$n_{\downarrow}$ in the ground state, it was shown in \cite{Fendley02} 
that for the closed chain of $N-2$ sites, $n_0$
is the integer closest to $(N-2)/3$.  By applying the
decomposition $Q=Q_1+Q_2$ discussed in \cite{Fendley02}, it is not
difficult to show that this applies to the open chain as
well. Basically, one proves that all states annihilated by $Q$ with
$n_{\downarrow}<n_0$ are not annihilated by $Q^\dagger$. Likewise, one
proves that all states annihilated by $Q^\dagger$ with
$n_{\downarrow}>n_0$ are not annihilated by $Q$. For the values of $N$ with
non-vanishing $W_N$, this means that
the ground state must have $n_{\downarrow}=n_0$.

We can now apply these results to the XXZ model
(\ref{eq:xxzsusyh}). For $N=3n_0+1$ for any integer
$n_0$, $W_N$ is non-zero. The integer closest to $3n_0-1$ is $n_0$,
so there must be a ground state at
$n_{\downarrow}=n_0$. This corresponds to
$L=N-n_{\downarrow}-1=2n_0$.
For $N=3n_0+2$, there is a ground state at
$n_{\downarrow}=n_0$, corresponding to $L=2n_0+1$.
Since the boundary magnetic field breaks the
$\uparrow\leftrightarrow\downarrow$ symmetry, it does not follow that
there is a second ground state for an odd length $L$. In fact, since the
boundary magnetic fields favors up spins, it is required that the ground
state for odd $L$ have more up spins than down spins.

To summarize, we have shown that the open XXZ chain with $\Delta=-1/2$
and boundary magnetic field $h=1/2$ has ground-state energy
$-3(L-1)/4$ for any $L$.  For an even number of sites $L$, the number
of down spins in the ground state is $n_\downarrow =L/2$; for odd $L$,
the ground state has $n_\downarrow=(L-1)/2$.  With this knowledge, one
should be able to extend the analysis of
\cite{FSZ,Stroganov,twistedXXZ} to find the polynomial describing the
ground state of this XXZ chain with a boundary magnetic field. We note
that our boundary magnetic field is different from the one utilized in
\cite{PS,FSZ}. There the field is complex (it is arises from demanding
quantum-group symmetry), and the Hamiltonian is not hermitian
(although its eigenvalues are real). The XXZ model with our real
boundary magnetic can be solved using the Bethe ansatz \cite{Gier2}.

Defining another charge $\widetilde{Q}$ by exchanging all up spins and
down spins in $Q$, then due to the obvious symmetry,
$\widetilde{Q}^2=0$. The Hamiltonian formed by $\widetilde{H}=
\{\widetilde{Q},\widetilde{Q}^\dagger\}$ is therefore also
supersymmetric. This yields the same XXZ Hamiltonian except with
boundary magnetic field $h=-1/2$, and so has the same ground-state
energy $(3L-1)/4$ as well.  One can then recover any magnetic field
with $|h|\le 1/2$ at the open ends by taking from a linear combination
of $H$ and $\widetilde{H}$. This proves that the ground-state energy
of the open chain is bounded from below by $-(3L-1)/4$ for any $|h|\le
1/2$, including the case of no magnetic field.  However, since $Q$ and
$\widetilde{Q}$ do not anticommute, this does not in general result in
a supersymmetric Hamiltonian, and the techniques utilized above do not
apply. Thus one can not prove using supersymmetry techniques whether
or not the bound is saturated for general $h$ like it is for $h=\pm
1/2$. It is easy to check for $L=3$ and $L=4$ that it is not.

\subsection{Periodic chain}

Since the supercharge defined above changes the total number of the
spins, it is natural to work on an open spin chain. Nevertheless, by
mapping the periodic XXZ chain with $\Delta=-1/2$ to the periodic $M_{1}$
chain\cite{Fendley03}, we can make some general statements about the
ground state of the XXZ model. In particular, we derive the
ground-state energy when the number of sites is odd.

The correspondence goes as follows \cite{Fendley03}.  Consider a state
$|S\rangle$ in the XXZ model with $L$ sites, and construct an
eigenstate of the translation operator $T$ by $|S_t\rangle\equiv
\big|S\rangle +t^{-1}T\big|S\rangle +
...+t^{-(L-1)}T^{L-1}\big|S\rangle$ for some root of unity $t$. Then
$|S_t\rangle$ is an eigenstate of $T$ with eigenvalue $t$, provided
that
\begin{equation}
t^{-(L-1)}T^{L}\big|S\rangle = t\big|S\rangle
\end{equation}
These are twisted boundary conditions when $t^L\ne 1$.  Let
$|S\rangle$ have $n_{\downarrow}$ down spins, and let $m_i\ge 1$ be
the distance between successive down spins. Then we can characterize
the state $|S_t\rangle$ by a series of $n_\downarrow$ integers
$m_1,m_2\cdots m_{n_\downarrow}$, with
$\sum_{j=1}^{n_\downarrow} m_j=L$. This characterization 
of the translationally-invariant eigenstates is one-to-one if we
identify cyclic permutations of the $m_i$.
Now we do likewise for the model $M_1$ on $N$ sites. 
For $f$ fermions one gets a set of integers $l_1,l_2,\dots,l_f$
with $\sum_{j=1}^f l_j =N$. Here the constraint is that $l_j\ge 2$, since
nearest-neighbor fermions are forbidden in the model $M_1$.
Thus there is a one-to-one map of translationally-invariant eigenstates,
if we make the identification $l_i = m_i+1$, $f=n_{\downarrow}$, 
and $N=L+n_{\downarrow}$. 
The one catch is that if we demand periodic boundary conditions
in $M_1$, we have $t^N=1$. 
This means that we must require twisted boundary conditions in the XXZ model:
\begin{eqnarray}
\label{eq:twistedbc}
S^{+}_{L+j} = tS^{+}_{j}\ ;\qquad\qquad
S^{-}_{L+j} = t^{-1}S^{-}_{j}\ .
\end{eqnarray}
Thus $t^N=t^{L+n_{\downarrow}}=1$ instead of the usual $t^L=1$.

With this mapping of states, one can then map the Hamiltonian of
the $M_1$ model to one acting on translationally-invariant eigenstates
of the XXZ model.
One finds \cite{Fendley03}
\begin{equation}
H_{XXZ}(\hbox{twisted})= H_{M_1}(\hbox{periodic}) - \frac{3L}{4}\ ,
\end{equation}
where the twisted boundary condition is (\ref{eq:twistedbc}).
For periodic boundary conditions in $M_1$, the Witten index is always
non-vanishing, so the lowest eigenvalue of $H_{M_1}$ is zero.  This
means the lowest eigenvalue in the corresponding sector of the twisted
XXZ model is $-3L/4$. Notice, however, that different translation
eigenvalues correspond to different boundary conditions. Thus when
this map is reversed, one finds XXZ states with periodic boundary
conditions can be mapped onto the $M_1$ model with twisted boundary
conditions, again depending on the translation
eigenvalue. Unfortunately, twisted boundary conditions in $M_1$ break
the supersymmetry in general, so one can no longer bound the
Hamiltonian. The exception is when we look for eigenstates of the
$XXZ$ model with periodic boundary conditions in the $t=1$
sector. These are mapped to the $M_1$ model with periodic boundary
conditions as well.

This means that supersymmetry is present in the periodic XXZ chain
only when $t=1$. Luckily, the ground state of the model
is in this sector. This follows from a proof that in a sector with fixed
$L$ and $n_{\downarrow}$, the ground state of the periodic XXZ model
is unique \cite{YY}.  This state must have $t=\pm 1$, because
otherwise there would be a second ground state with eigenvalue
$t^{-1}$. With the choice of sign of the $S^x S^x$ and $S^y S^y$ terms
in (\ref{eq:xxz-Hamiltonian}), the ground state clearly has $t=1$.
(Besides, for an odd number of sites, the eigenvalue $t=-1$ is not
allowed for periodic boundary conditions.)  This is not surprising,
given that when writing the Hamiltonian acting on eigenstates of $T$,
the entries in the Hamiltonian are the most negative for $t=1$.

The results of \cite{Fendley02,Fendley03} therefore give the
ground-state energy of the XXZ model with periodic boundary conditions
when $L$ is odd. When $N$ is an not a multiple of $3$, the number of
fermions in the ground state $f_0$ is the integer closest to $N/3$,
and the translation eigenvalue is $t=1$. Thus when $N=3f_0+1$, the map
to XXZ takes this to $L=N-f_0 = 2f_0+1 = 2n_{\downarrow}+1$.  We thus
know the ground state for an odd number of sites has
energy $-3L/4$. When $N=3f_0-1$, the map to XXZ takes this to a state
with $L=2n_{\downarrow}-1$. Thus we recover both ground states of 
the periodic XXZ chain with an odd number of sites.

When $N$ is a multiple of $3$, there are two zero-energy eigenstates
of $M_1$, which have $t\ne 1$. This state maps to an XXZ model with an
even number of sites, but with twisted boundary conditions. It
therefore says nothing about the XXZ model with periodic boundary
conditions, but does imply we can find the ground-state energy with
twisted boundary conditions. Thus our results are in harmony with the
analysis of \cite{Stroganov,twistedXXZ}. This work shows that the
roots of the Bethe equations of the XXZ model at $\Delta=-1/2$ are
given in terms of a single polynomial equation, when there are an odd
number of sites and periodic boundary conditions, or an even number of
sites and twisted boundary conditions. These are the supersymmetric
cases!

\section{Supersymmetry in the t-J model}

Another model, denoted $M_{2}$, was explored in depth in
\cite{Fendley03}.  It is a model of spinless fermions on a chain with
the constraint that no more than two consecutive sites can be
occupied. We can thus think of this, roughly, as a three-state system:
empty sites, lone occupied sites, and nearest-neighbor occupied sites.
This correspondence was used in \cite{Fendley03} to map $M_2$
to two different familiar three-state
models. At one coupling, it can be mapped to
the spin-1 generalization of the XXZ chain. By repeating the above analysis
to find the exact ground-state energy, one presumably could find
to find the single polynomial equation the corresponding Bethe roots obey.

At a value of the coupling where the Hamiltonian preserves the number
of each of these three states, $M_2$ was mapped to the ferromagnetic
t-J model.  In particular if an edge between empty sites in the $M_2$
model is mapped to an up-spin, a lone occupied site to a down-spin,
and an edge between two adjacent occupied sites to a hole, then the
model is related to the t-J model in the same fashion as the $M_1$ is
related to the spin-$1/2$ XXZ chain.  As with the XXZ model, the
spectrum of the two was shown to be the same, when certain twisted
boundary conditions are utilized. This therefore hints that like the
XXZ model, the supersymmetry can be realized directly in this t-J
model at this special point. In this section we show that this is so:
there exists a non-local space-time supersymmetry for the t-J model in
an enlarged Hilbert space. As a byproduct, we also explicitly show
that the $M_{2}$ lattice model has an SU(2) symmetry as well.

The one-dimensional t-J model describes fermions with spin hopping
along a chain.  Double occupancy is forbidden, so it is convenient to
think of this as a three-state system, with an empty site being
created by a bosonic operator $b_i^\dagger$. The Hamiltonian is
\begin{eqnarray} \label{eq:t-J-Hamiltonian}  H &=&
-t\sum_{i,\sigma}\left(d^{\dagger}_{i\sigma}d_{i+1,\sigma} +
h.c\right) +2\sum_{i} n_{i} -
L\\ \nonumber & &+ J\sum_{i}\left[S^{z}_{i}S^{z}_{i+1}
+ \frac{1}{2}\left(S^{\dagger}_{i}S_{i+1}^{}+ S^{\dagger}_{i+1}S_{i}^{}\right)
- \frac{1}{4}n_{i}n_{i+1}\right] 
\end{eqnarray}
where $f_{i\sigma}$ ($\sigma = \uparrow, \downarrow$) annihilates an
fermion, so that the composite operator
$d_{i\sigma} = b^{\dagger}_{i}f_{i\sigma}$ removes a fermion and 
creates a boson.
The usage of the composite operator ensures no doubly occupied
states will be generated. The $n_{i}$ are the 
fermion number operators and the $S_{i}$ are
spin operators:
$$
n_{i\sigma} = f^{\dagger}_{i\sigma}f_{i\sigma}$$
$$S^{z}_{i} =
\frac{1}{2}\left(f^{\dagger}_{i\uparrow}f_{i\uparrow}-f^{\dagger}_{i\downarrow}f_{i\downarrow}\right)\qquad
S^{\dagger}_{i} = f^{\dagger}_{i\uparrow}f_{i\downarrow}\qquad
S_{i} - f^{\dagger}_{i\downarrow}f_{i\uparrow}$$

At $J=\pm 2t$, and the particular chemical potential given in
(\ref{eq:t-J-Hamiltonian}), the t-J Hamiltonian has a global $u(1|2)$
symmetry rotating the three states on each site into each other
\cite{Sarkar}. One can thus think of the $J=2t$ case as the
antiferromagnetic $u(1|2)$ Heisenberg model, and the $J=-2t$ case as the
ferromagnetic one. Three of the nine generators of $u(1|2)$ symmetry
are fermionic, while the other six are bosonic, so this is a graded
Lie algebra. The generators are
\begin{eqnarray*}
\label{eq:9gen}
J_{i,1} = S^{+}_{i} = f^{\dagger}_{i\uparrow}f_{i\downarrow}\ ,\qquad
J_{i,2} &= S^{-}_{i} = f^{\dagger}_{i\downarrow}f_{i\uparrow}\ ,\qquad
J_{i,3} &= S^{z}_{i} = \frac{1}{2}(n_{i\uparrow}-n_{i\downarrow})\\ \nonumber
J_{i,4} = (1-n_{i,\downarrow})f_{i\uparrow}\ ,\qquad
J_{i,5} &= J_{i,4}^{\dagger}\ ,\qquad\qquad\qquad
J_{i,6} &= (1-n_{i,\uparrow})f_{i\downarrow}\\ \nonumber
J_{i,7} = J_{i,6}^{\dagger}\ ,\qquad
J_{i,8} &= 1-\frac{1}{2}n_{i}\ ,\qquad\qquad
J_{i,9} &= 1
\end{eqnarray*}
Because of the fermionic generators, such a
symmetry is often called ``supersymmetry'' in condensed matter
physics. One should keep in mind that it is different from the
space-time supersymmetry used above and to be used below. In both
cases the Hamiltonian commutes with the symmetry algebra, but for
space-time supersymmetry, the Hamiltonian (\ref{eq:anticom})
is a non-trivial part of the algebra.

We find that in the ferromagnetic case $J=-2t$ 
that, in addition to the graded Lie algebra, there exists a
space-time supersymmetry whose generators are given by
$Q^{(\dagger)}=\sum_{i}Q^{(\dagger)}_{i}$ with \begin{equation}
\label{eq:t-J-supercharge}  Q_{i} =
A_{i}K_{i}\left(f_{i+1,\uparrow}d_{i\downarrow} -
f_{i+1,\downarrow}d_{i\uparrow}\right)
\end{equation}
As in to the last section, $A_{i}$ shifts all the sites $j$ with
$j>i+1$ to the $j-1$th sites, while $K_{i}=(-1)^{\sum_{j<i}n_{h}}$ is
a ``string'' depending on the number of holes to the left of site i.
As in a Jordan-Wigner transformation, this string is what makes $Q$
fermionic. Both the shift operator $A$ and the string $K$ make $Q$ act
non-locally. Physically, $Q_{i}$ annihilates a pair of electrons with
opposite spins at site $i$ and $i+1$, creates a hole at site $i$, then
shift all the sites to the right of $i$ to the left by one. There are
two things worth mentioning about the supercharges.

Let us give a little more detail. We first consider
$\{Q, Q\}$. It is easy to see due to the $K_{i}$ factors,
$A_{i}K_{i}f_{i+1,\uparrow}d_{i\downarrow}A_{j}K_{j}f_{j+1,\uparrow}d_{j\downarrow}$
is canceled by
$A_{j}K_{j}f_{j+1,\uparrow}d_{j\downarrow}A_{i}K_{i}f_{i+1,\uparrow}d_{i\downarrow}$.
Similar cancellations occur for the other three terms of the
anticommutator, so $Q$ and $Q^{\dagger}$ are nilpotent. For 
$H=\{Q, Q^{\dagger}\}$, we have
\begin{eqnarray}
\label{eq:mainanti} \{Q,Q^{\dagger}\} &=&
\sum_{i,j}\{A_{i}K_{i}f_{i+1,\uparrow}d_{i\downarrow},
d^{\dagger}_{j\downarrow}f^{\dagger}_{j+1,\uparrow}K_{j}A^{\dagger}_{j}\}
-\{A_{i}K_{i}f_{i+1,\uparrow}d_{i\downarrow},
d^{\dagger}_{j\uparrow}f^{\dagger}_{j+1,\downarrow}K_{j}A^{\dagger}_{j}\}\\
\nonumber & &-\{A_{i}K_{i}f_{i+1,\downarrow}d_{i\uparrow},
d^{\dagger}_{j\downarrow}f^{\dagger}_{j+1,\uparrow}K_{j}A^{\dagger}_{j}\}
+\{A_{i}K_{i}f_{i+1,\downarrow}d_{i\uparrow},
d^{\dagger}_{j\uparrow}f^{\dagger}_{j+1,\downarrow}K_{j}A^{\dagger}_{j}\}
\end{eqnarray}
After the cancellations due to the $K_{i}$ factors,
we have the following terms:\newline
(1)$-A_{i}K_{i}f_{i+1,\uparrow}d_{i,\downarrow}d^{\dagger}_{i+1,\uparrow}f^{\dagger}_{i+2,\downarrow}K_{i+1}A^{\dagger}_{i+1}$ 
and its Hermitian conjugate give us the down spin hopping term, exchanging up
and down spins, we get the up spin hopping term. \newline
(2)$-d^{\dagger}_{i,\uparrow}f^{\dagger}_{i+1,\downarrow}K_{i}A^{\dagger}_{i}A_{i}K_{i}f_{i+1,\uparrow}d_{i,\downarrow}$ 
and its Hermitian conjugate gives us neighboring opposite spins exchange term.
\newline
(3)$d^{\dagger}_{i,\uparrow}f^{\dagger}_{i+1,\downarrow}K_{i}A^{\dagger}_{i}A_{i}K_{i}f_{i+1,\downarrow}d_{i,\uparrow}$ 
and its Hermitian conjugate count the number of bonds between opposite
spins.\newline
(4)$A_{i}K_{i}f_{i+1,\uparrow}d_{i,\downarrow}d^{\dagger}_{i,\uparrow}f^{\dagger}_{i+1,\downarrow}K_{i}A^{\dagger}_{i}$  counts the number of holes. \newline
These terms yield the Hamiltonian
\begin{eqnarray}
\label{eq:finalH2}
 H =
-\sum_{i,\sigma}\left(d^{\dagger}_{i\sigma}d_{i+1,\sigma} +
h.c\right)-2\sum_{i}\left[S^{z}_{i}S^{z}_{i+1} +
\frac{1}{2}\left(S^{\dagger}_{i}S_{i+1}+ h.c\right) -
\frac{1}{4}n_{i}n_{i+1}\right] +2N_{h}
\end{eqnarray}
where $N_{h}$ is the total number of holes. Since $N_h = L - f$, where
$L$ is the length and $f$ is the number of fermions, we recover the
ferromagnetic t-J Hamiltonian (\ref{eq:t-J-Hamiltonian}) up to a shift
$L$. The ground states of both the $M_2$ model and this ferromagnetic
t-J model are discussed in \cite{Fendley03}.

We can explore the symmetry structure of the t-J model a little more by
considering the commutators of the supercharge and the $SU(2)$ generators,
namely, $S^{+}, S^{-}$ and $S^{z}$. Obviously, $Q$ commutes with $S^{z}$,
while the commutator of $Q$ and the other two give us $Q^{(1)}$, which is
very similar to $Q$, but it annihilates a pair of electrons with same
instead of opposite spins. It is easy to see the square of $Q^{(1)}$ also
vanishes, and $Q^{(1)}$ also commutes with the t-J Hamiltonian, but it does
not generate the Hamiltonian.

We have shown that the ferromagnetic t-J model has supersymmetry. It is
also interesting to ask if the $SU(2)$ symmetry exists in the $M_{2}$
supersymmetric fermion model. The answer is yes: 
the counterpart of $S^{+}$ is the operator which annihilates
a single fermion on site $i$ and shifts all the sites $j>i$ to the left
by one. Since this operation is fermionic, in order to
"defermionize" it, we need a factor $(-1)^{\sum_{j<i}n_{j}}$, which
depends on the number of fermions to the left of site $i$. So we see
that the supersymmetry and $SU(2)$ symmetry exist in both the t-J 
and $M_{2}$ models.  The supercharge in the $M_{2}$ model is local, but
it is non-local in the t-J model, while the local $SU(2)$ generators in
the t-J model become non-local in the $M_{2}$ model.

\section{Generalization to $R^4=0$}

With appropriate boundary conditions, the XXZ model on an open chain
is known to have a quantum-group symmetry for all $\Delta$
\cite{PS}. The simplest non-trivial example is at $\Delta=-1/2$, where
the quantum group generators $S^{\pm}$ are nilpotent:
$(S^{\pm})^2=0$. Given the supersymmetry algebra discussed above, this
hardly can be a coincidence. However, the two symmetries are not
identical: for example, the boundary conditions required for the
quantum-group symmetry result in imaginary boundary magnetic fields;
such fields obviously cannot be obtained from a hermitian Hamiltonian
like ours.

The quantum-group symmetry exists at any value of $\Delta$.
For example, when $\Delta=-\cos(\pi /s)$ for integer $s$, the
quantum group generators obey $(S^{\pm})^{s-1} =0$. 
The similarity of
supersymmetry and the quantum-group symmetry at $\Delta =-1/2$ led us
to attempt to construct XXZ-type lattice models obeying similar
nilpotency relations.   A similar idea was pursued in \cite{Leclair}, in
the context of the particle description of sine-Gordon field theory.

We consider a generalization of our method to a model with $R^4=0$.
We study a lattice model with same Hilbert space as $M_1$: spinless
fermions $c_j$ forbidden to be adjacent. If we take
\begin{equation} 
R_{j}=P_{j+1}P_{j-1}c_{j}\exp\left[-i\pi\sum_{l<j}n_{l}/2\right]
\end{equation}
where $P_{j}=1-c^\dagger_{j}c_j$ is a projection operator.
A Hamiltonian which commutes with the charge $R=\sum_j R_j$ has been
constructed \cite{Leclair}; it is
\begin{equation}
H= [R^{\dagger}, R]^{2} -R^{\dagger^{2}}R^{2} -R^{2}R^{\dagger^{2}}
\end{equation}
There are two nice equalities satisfied by $R$:
\begin{eqnarray}
R^{4}&=&0\\
\{R^{\dagger^{2}}, [R^{\dagger}, R]\}&=&0
\end{eqnarray}
They can be easily checked by, again, keeping track of the
configuration change. The Hamiltonian includes the following terms,
translated into XXZ language by the same arguments as given above.
The first one is the next-nearest-neighbor hopping term with the
coefficient 1; it interchanges
$$\downarrow\;\;\;\uparrow\;\;\;\uparrow \quad\leftrightarrow \quad
\uparrow\;\;\;\uparrow\;\;\;\downarrow$$
The second term is the nearest neighbor hopping term, which in XXZ
language changes
\begin{eqnarray*}
\downarrow\;\;\;\downarrow\;\;\;\uparrow\;\;\;\downarrow 
&\quad\rightarrow\quad&
\downarrow\;\;\;\uparrow\;\;\;\downarrow\;\;\;\downarrow
\quad\hbox{ with magnitude }2;\\
\uparrow\;\;\;\downarrow\;\;\;\uparrow\;\;\;\uparrow &\quad\rightarrow\quad&
\uparrow\;\;\;\uparrow\;\;\;\downarrow\;\;\;\uparrow 
\quad\hbox{ with magnitude }2-2i;\\
\uparrow\;\;\;\downarrow\;\;\;\uparrow\;\;\;\downarrow &\quad\rightarrow\quad&
\uparrow\;\;\;\uparrow\;\;\;\downarrow\;\;\;\downarrow
\quad\hbox{ with magnitude }2-i;\\
\downarrow\;\;\;\downarrow\;\;\;\uparrow\;\;\;\uparrow &\quad\rightarrow\quad&
\downarrow\;\;\;\uparrow\;\;\;\downarrow\;\;\;\uparrow
\quad\hbox{ with magnitude }2-i,
\end{eqnarray*}
plus all the Hermitian conjugates. The third term is the potential
term, for XXZ model; it counts the number of adjacent up spins and single
down spins, and also assigns the potentials to the following
configurations: $\uparrow\;\;\downarrow\;\;\uparrow$ with
coefficient $2$; $\uparrow\;\;\downarrow\;\;\downarrow$ with
coefficient $1$; $\downarrow\;\;\downarrow\;\;\uparrow$ with
coefficient $1$; and $\downarrow\;\;\downarrow\;\;\downarrow$ with
coefficient $0$. All these terms look unsymmetric, in order to make it
look more symmetric and elegant, we add terms generated by another charge
which interchanges down and up spins in our first charge and changes $i$ to
$-i$. After adding two Hamiltonians together, we get
\begin{equation}
H= \sum_{j} \{S^{+}_{j}S^{-}_{j+2} +(4-2i)S^{+}_{j}S^{-}_{j+1} +h.c
-\frac{1}{2}S^{z}_{j}S^{z}_{j+1}\}
\end{equation}
This is a generalized XXZ chain which involves next-nearest-neighbor
interactions which exist in the $xy$ plane, but not in the $z$
direction. So if we rotate spins around $z$ axis by certain angles
which depend on the site index, the coefficient of this term can be
made purely imaginary, and thus can be taken away by adding another
term with $i$ replaced by $-i$. Unfortunately, to get this Hamiltonian
we needed to add two different Hamiltonians (as in the open XXZ chain
with variable boundary field). But since each charge does not commute
with sum of the two Hamiltonian, the symmetry does not seem to
persist.  Thus we are not sure if the above generalization will shed
any light on the analysis of XXZ model at other values of $\Delta$.

\section{Conclusion}

In this paper, we explicitly constructed the supercharges which
generate space-time supersymmetries for XXZ and t-J chains. We find
that an open XXZ chain at $\Delta =-\frac{1}{2}$ with a particular
magnetic field at both ends is supersymmetric. The supercharges change
the total number of sites $L$ of the spin chain, while conserving the
quantity $L+n_{\downarrow}$. This enabled us to find that the
ground-state energy for any $L$ in the magnetic field ensuring
supersymmetry, and bound its value for any smaller field.
For periodic boundary conditions, we find the exact ground-state
energy for an odd number of sites. For an even number of sites, 
the supersymmetry only survives for twisted boundary conditions.
These particular boundary conditions are precisely the cases where
the Bethe equations for the XXZ model ground state can be simplified
dramatically \cite{Stroganov,twistedXXZ}.  

We also showed space-time
supersymmetry and global super Lie algebra $u(1|2)$ coexist in the
ferromagnetic t-J model at $2t=-J=2$. Since the supercharges can be
constructed out of the generators of $u(1|2)$, the coexistence of the
two symmetries might indicate some intimate relations between them,
further work is required to uncover the hidden relation.

A common feature of these supercharges is the non-locality. Non-local
symmetries arise in a number of interesting two-dimensional classical
and one-dimensional quantum systems as hidden symmetries. A
famous example is the Yangian symmetry of the $O(3)$ sigma model
\cite{Luscher}. We hope the supersymmetry discussed in this paper
will improve our understanding of such models, and provide new insights
into other problems.

\bigskip\bigskip

We would like to thank Jan de Gier, Bruno Nachtergaele and Chetan
Nayak for useful correspondence.  P.F.\ would also like to thank
Kareljan Schoutens, Jan de Boer and Bernard Nienhuis for many
conversations and collaboration on \cite{Fendley02,Fendley03}. This
work was supported by the National Science Foundation through the
grant NSF-DMR-0104799.  The work of P.F.\ was also supported by the
DOE under grant DEFG02-97ER41027.

\end{document}